\documentstyle[psfig]{mn}
  \title[Hypercritically Accreting Neutron Stars]{Influence of the
         r-mode instability on hypercritically accreting neutron stars}
  \author[Yoshida \& Eriguchi]{Shin'ichirou~Yoshida
    \thanks{yoshida@sissa.it}
    \thanks{present address: Scuola Internazionale Superiore di Studi Avanzati, Via Beirut 2-4, 34014 Trieste, Italy}
          and Yoshiharu~Eriguchi\thanks{eriguchi@valis.c.u-tokyo.ac.jp}\\
          Department of Earth Science and Astronomy, 
          Graduate School of Arts and Sciences, 
          University of Tokyo,\\
          Komaba, Meguro-ku, Tokyo 153-8902, Japan}
   \date{Accepted, Received}
\begin{document}
\maketitle
%
\begin{abstract}
%
We have investigated an influence of the r-mode instability on 
hypercritically accreting ($\dot{M}\sim 1M_\odot \mbox{y}^{-1}$) neutron 
stars in close binary systems during their common envelope phases
based on the scenario proposed by Bethe et al.~\shortcite{bethe-brown-lee}.
On the one hand neutron stars are heated by the accreted matter at the 
stellar surface, but on the other hand they are also cooled down by 
the neutrino radiation. At the same time, the accreted matter transports 
its angular momentum and mass to the star. We have studied the evolution of 
the stellar mass, temperature and rotational frequency. 

The gravitational-wave-driven instability of the r-mode oscillation strongly 
suppresses spinning-up of the star, whose final rotational frequency is well
below the mass-shedding limit, typically as small as $10\%$ of that of the 
mass-shedding state. On a very short time scale the rotational frequency 
tends to approach a certain constant value and saturates there as far as the 
amount of the accreted mass does not exceed a certain limit to collapse to 
a black hole. This implies that the similar mechanism of gravitational 
radiation as the so-called Wagoner star may work in this process. The star 
is spun up by accretion until the angular momentum loss by gravitational 
radiation balances the accretion torque. 
The time-integrated dimensionless strain of the radiated gravitational wave 
may be large enough to be detectable by the gravitational wave detectors 
such as LIGO II.
%
\end{abstract}
%


\section{Introduction}

In relation to the formation mechanism of compact binary systems such as double
neutron star systems, double black hole systems or black hole--neutron star 
systems, hypercritical accretion flows onto neutron stars in envelopes of 
massive stars of close binary systems have been investigated by several 
authors~\cite{chevalier93,chevalier96,brown,bethe-brown,bethe-brown-lee}. 
The mass accretion rate ($\dot{M}$) of the hypercritical accretion flow 
amounts to $10^8$ times as high as the Eddington's mass accretion limit.
Chevalier~\shortcite{chevalier96} found that for rather
small values of the viscosity parameter, $\alpha\sim 10^{-6}$, the accretion 
flow which cools via neutrino radiation could be realized. 

Recently Bethe et al.~\shortcite{bethe-brown-lee} pointed out that the flow 
with much larger and conservative values of the viscosity parameter, 
$\alpha\ga 0.1$, does not become so hot as to emit neutrinos  
during the accretion process but that the accretion will proceed to the 
advection dominated accretion flow (ADAF). In their model, the accreted matter 
onto the neutron star surface is processed there by nuclear reactions and 
cools via neutrino emission. The hypercritical accretion increases the mass of 
the star rapidly and induces the collapse of the neutron star to a black hole.
Thus the supernova explosion of the massive companion will lead to the 
formation of a black hole--neutron star binary system or a double black hole
binary system. The merging event of such a system might be one of the 
important targets for the gravitational wave astronomy. 

In the accretion process, a non-negligible amount of angular momentum 
is carried into the neutron star. This potentially could spin up the 
star to the Kepler limit, i.e. to the state where the stellar rotational 
frequency equals to the orbital frequency at the equatorial surface. The star
rotating at the Kepler limit deforms significantly from the spherical
configuration and the subsequent collapse of the rapidly rotating star to a 
black hole might radiate a substantial fraction of the binding energy of the 
system in the form of gravitational waves. 

This seemingly straightforward conclusion, however, has to be reexamined
from the standpoint of stability of rotating neutron stars. The gravitational 
radiation driven instability of rotating stars might prevent the stellar 
rotational frequency to increase above a certain limit. In particular, 
the recently discovered r-mode instability~\cite{andersson,friedman-morsink} 
might be sufficiently strong to suppress the spin frequency in some range of 
the stellar temperature.

In this paper, in order to see how this instability influences the accretion 
induced spin-up of the neutron star in the hypercritical accretion, we will
employ the hypercritical ADAF model based on the analysis of Bethe et 
al.~\shortcite{bethe-brown-lee} and investigate the evolution of the 
temperature and the rotational frequency of the neutron star.
\footnote{Similar investigations of accreting neutron stars 
with much lower accretion rate, which are the models of low mass X-ray 
binaries, can be seen for instance 
in Andersson et al.~\shortcite{andersson-kokkotas-stergioulas} 
and in Levin~\shortcite{levin}.}

\section{Simplified Model of the Accretion Induced Evolution of Neutron Stars}

It should be noted that the study in this paper is done in the
framework of Newtonian dynamics and Newtonian gravity. The gravitational 
radiation reaction on stellar oscillations is evaluated by the lowest order 
post-Newtonian approximation~\cite{thorne}. Furthermore to simplify our
analysis, stellar configurations are assumed to be spherical polytropes,
$p=K\rho^2$, with the polytropic index $N=1$. 
Here $p$ and $\rho$ are the pressure and the density, respectively.
By choosing the polytropic coefficient $K$ appropriately, the stellar mass
is $M=1.4M_{\odot}$ and its radius is $R=12.5$km. Note that the radius of
the polytropes with index $N=1$ is independent of the central density
and thus of the mass if the constant $K$ is fixed.

\subsection{Characteristics of the hypercritical ADAF onto neutron stars}

The hypercritical accretion flow which we employ in this paper is proposed 
by Bethe et al.~\shortcite{bethe-brown-lee}. The 
characteristics of that flow are summarized as follows. 
(1) The flow is hypercritical because the accretion rate is typically 
$\sim 1 M_{\odot}\mbox{y}^{-1}$.
(2) Since the flow is advection dominated~\cite{narayan-yi} and its
temperature is relatively low ($\sim 0.5$MeV), the adiabatic 
exponent of the gas equals to $4/3$ (radiation dominated mixture of 
photon, electron, positron and nuclei).
(3) The accreted matter touches down on the surface of the star `softly' and
burns to produce as heavy nuclei as $^{56}\mbox{Ni}$. On the other hand, 
the matter is cooled by neutrino emission. Consequently the whole process 
acts as a `thermostat' which keeps the neutrino temperature $\sim 1$MeV.

\subsection{Mass accretion rate}

The mass accretion rate $\dot{M}_{\rm acc}$, which is the rate of mass 
attachment to the star, is assumed to be constant during the evolution. In 
general the mass accretion rate should be distinguished from the mass inflow 
rate $\dot{M}_{\rm in}$, which is the total amount of the incoming mass per 
unit time to the neutron star. This is because some kinds of outflow may 
coexist with the accreting flows. In this paper we simplify the situation by 
assuming that both rates are the same, i.e. $\dot{M}_{\rm in}=\dot{M}_{\rm acc}
\equiv\dot{M}$. We will discuss this assumption further in {\it Discussion}. 

The evolution of the neutron star mass $M$ is written as:
\begin{equation}
	\frac{M}{M_{\odot}} = \mu_0 + \beta t \ ,
\label{eq-mass}
\end{equation}
where $\mu_0$ is the initial mass of the star (in units of the solar mass), 
time $t$ is measured in units of second, and $\beta$ is defined as:
\begin{equation}
	\beta = 3.2\times 10^{-8} \left[\frac{\dot{M}}{1M_{\odot}\mbox{y}^{-1}}\right] \ .
\end{equation}

\subsection{Thermal response of the neutron star heated by neutrino from 
the surface}

The accreted matter landing on the surface layer is processed 
by nuclear 
burning and emits neutrinos. We will assume that roughly a half of them 
escapes from the star freely and that some fraction of the rest of them 
interacts with stellar matter and heats it up. The mean free path $\lambda$ of 
inelastic scattering of neutrinos with electrons in the neutron star matter 
is written as~\cite{shapiro-teukolsky},
\begin{equation}
    \lambda \sim 2\times10^{10} \left[\frac{\rho_{\rm nuc}}{\bar{\rho}}
    \right]^{7/6} \left[\frac{0.1 \mbox{MeV}}{E_\nu}\right]^{5/2} \mbox{cm} \ ,
\end{equation}
where $\rho_{\rm nuc}$ is the `nuclear density' ($\sim 2.8\times 10^{14} 
\mbox{g cm}^{-3}$) and $E_{\nu}$ is the neutrino energy. The density 
$\bar{\rho}$ means the averaged density of the star, which is related in our 
polytropic stellar models to the mass $M$ as follows:
\begin{equation}
	\frac{\bar{\rho}}{\rho_{\rm nuc}} = 0.7 \frac{M}{M_{\odot}}  \ .
\end{equation}
The neutrino energy $E_{\nu}$ is fixed to $1$MeV. We simply assume the 
scattering (thus heating) rate as $\eta R/\lambda$ where $R$ is the stellar 
radius and $\eta$ is an efficiency factor of $\nu$-heating through the 
inelastic scattering.
As the flow is adiabatic and radiation dominated, the specific internal 
energy of the accreted matter $\varepsilon$ is written as,
\begin{equation}
	\varepsilon = \frac{3}{4}\times\frac{6}{7}\frac{GM}{R} \ .
\end{equation}
With these expressions, the heating rate of the whole star 
$\dot{\varepsilon_{\rm h}}$ is,
\begin{eqnarray} 
	\dot{\varepsilon_{\rm h}} &\sim& \dot{M} \varepsilon \frac{\eta R}{\lambda}\nonumber\\
		&\sim& 4.7\times 10^{43} \left[\frac{\dot{M}}{1 M_{\odot} \mbox{y}^{-1}}\right]\left[\frac{M}{M_{\odot}}\right]^{13/6} \mbox{erg} \ \mbox{s}^{-1}.
\end{eqnarray}
Here we choose $\eta = 1$. Heating rate above is proportional to the 
multiplication of $\dot{M}$ and $\eta$. Thus the reduction of $\eta$
is equivalent to that of $\dot{M}$ with $\eta$ being fixed. 
As shown later in Discussion, the dependence of the results on $\eta$ is weak.
\footnote{Here heating of the star by the r-mode oscillation, which is considered in Levin~(1999), is not taken into account. We confirmed, however, that an inclusion of the r-mode heating term is negligible in the situation of our interest.}

As the cooling mechanism of the star, we assume 
the modified URCA process to be dominant~\cite{shapiro-teukolsky}. Its 
cooling rate $L_{\nu}^{\rm URCA}$ is,
\begin{equation}
	L_{\nu}^{\rm URCA} = 6\times 10^{39} \left[\frac{M}{M_{\odot}}\right]^{2/3} T_9^8 \ ,
\end{equation}
where $T_9$ is the stellar temperature, which is assumed to be uniform, in 
units of $10^9$K. The heat capacity $C_{\rm v}$ of the neutron star which is 
approximated by the following formula if the neutron star is assumed to 
consist of degenerate free nucleons,
\begin{equation}
	C_{\rm v} = 8\times 10^{54} k_B \times \left[\frac{M}{M_{\odot}}\right]\left[\frac{\bar{\rho}}{\rho_{\rm nuc}}\right]^{-2/3} T_9 \ ,
\end{equation}
where $k_B$ is the Boltzmann constant. Then the equation of the thermal 
balance of the star can be written as,
\begin{equation}
	\frac{d}{dt}\left(\frac{1}{2} C_{\rm v} T\right) = \dot{\varepsilon_{\rm h}} - L_{\nu}^{\rm URCA} \ .
\end{equation}
This is transformed to the equation of the evolution of the temperature
as follows:
\begin{eqnarray}
	\frac{d}{dt}\left(\left[\frac{M}{M_{\odot}}\right]^{1/3}T_9^2\right) &=& 6.5\times10^{-5} \left[\frac{\dot{M}}{1 M_{\odot}\mbox{y}^{-1}}\right]\left[\frac{M}{M_{\odot}}\right]^{13/6}\nonumber\\ 
&-& 8.7\times 10^{-9}\left[\frac{M}{M_{\odot}}\right]T_9^8 \ .
\label{eq-thermal}
\end{eqnarray}

\subsection{Angular momentum balance of the star}

The angular momentum of the star is changed by two torques induced
by accretion and by gravitational radiation if the r-mode oscillation is 
unstable.

The accretion torque is estimated as $\dot{M}j_a$, where $j_a$ is the specific
angular momentum of the accreted matter:
\begin{equation}
j_a = R \sqrt{\frac{2}{7} \frac{GM}{R}} \ .
\end{equation}
Here the ADAF solution by Narayan \& Yi~\shortcite{narayan-yi} is employed 
to evaluate the rotational velocity of the accreted matter at 
the stellar surface.

As for the gravitational radiation torque, we use the simple formula
used in Lindblom et al.~\shortcite{lindblom-owen-morsink} which also
incorporates the damping effect by viscosity. This corresponds to the 
assumption that the r-mode quickly grows to saturate in the non-linear regime 
of its amplitude and the angular momentum of the mode is a substantial 
fraction of the stellar angular momentum. 

Let us introduce the rate $\delta$ of the angular momentum of the mode to
the stellar angular momentum. By using the moment of inertia of the star $I$ 
and the stellar rotational angular frequency $\Omega$, the torque due to
gravitational wave emission can be written as 
$\tau_{\rm r}^{-1}I\Omega \delta$. The factor $\delta$ is related
to the dimensionless amplitude of the oscillation of the r-mode `$\kappa$',
 which is used in the simplified discussion of the non-linear evolution of 
the r-mode by Owen et al.~(1998), as $\delta \sim \kappa^2$. 

The time scale of the r-mode instability $\tau_{\rm r}$ is then defined as,

\begin{equation}
   \tau_{\rm r}^{-1} = \tau_{\rm gr}^{-1} + \tau_{\rm s}^{-1} 
        + \tau_{\rm b}^{-1}
\end{equation}
where $\tau_{\rm gr}$ is the energy dissipation time scale by the 
gravitational radiation, $\tau_{\rm s}$ is that of the shear viscosity due to 
neutron-neutron collision and $\tau_{\rm b}$ is that of the bulk viscosity
due to the lag of the $\beta-$reaction and the stellar oscillation.
The negative value of $\tau_{\rm r}$ signals the onset 
of the instability. We set $\tau_{\rm r}^{-1}$ to be zero when this value is 
positive, i.e. for damping oscillations.

The time scale of gravitational wave dissipation is obtained as,
\begin{eqnarray}
	\tau_{\rm gr}^{-1} &=& -\frac{32\pi G}{c^{2l+3}}\cdot\frac{(l-1)^{2l}}{[(2l+1)!!]^2}\left(\frac{l+2}{l+1}\right)^{2l+2}\nonumber\\
&&\times \Omega^{2l+2}\rho_c R^{2l+3}\int_0^1 \tilde{\rho}(x)x^{2l+2} dx.
\end{eqnarray}
Here $\tilde{\rho}$ is the density distribution normalized by the central
density $\rho_c$.
The central density is related to the stellar mass $M$ for $N = 1$ polytropes 
as,
\begin{equation}
	\rho_c = 3.29 \bar{\rho} = 3.29\times\rho_{\rm nuc}\times 0.7\frac{M}{M_{\odot}} \ .
\end{equation}
Thus for $l=m=2$ modes for which the instability is the strongest,
\begin{equation}
	\tau_{\rm gr}^{-1} = -3.1\times 10^{-8} \left[\frac{f}{100\mbox{Hz}}\right]^6 \left[\frac{M}{M_{\odot}}\right].
\end{equation}
Here $f$ is the rotational frequency of the star ($f = \Omega/2\pi$). 

The damping time scales due to shear and bulk viscosities are,
\begin{eqnarray}
	\tau_{\rm s}^{-1} &=& (l-1)(2l+1)\int_0^R\eta r^{2l}dr/\int_0^R\rho r^{2l+2} dr\nonumber\\
&=& 5\times 347\rho_c^{9/4}R^5 10^{-18}T_9^{-2}\int_0^1\bar{\rho}(x)^{9/4}x^4dx\nonumber\\
&&/\left(\rho_c R^7\int_0^1\bar{\rho}(x)x^6 dx\right)\nonumber\\
	&=& 2.0\times 10^{-9} \left(\frac{M}{M_{\odot}}\right)^{5/4} T_9^{-2},
\end{eqnarray}
and,
\begin{eqnarray}
	\tau_{\rm b}^{-1} &=& (6.99\times 10^8)^{-1}\left(\frac{2\pi f}{\pi G\bar{\rho}}\right)^2 T_9^6\nonumber\\
	&=& 1.4\times 10^{-11} \left[\frac{f}{100\mbox{Hz}}\right]^2 \left[\frac{M}{M_{\odot}}\right]^{-1} T_9^6.
\end{eqnarray}
%

The equation for the evolution of the rotational frequency is written as,
\begin{eqnarray}
	&&\frac{d}{dt}\left(\left[\frac{M}{M_{\odot}}\right]\left[\frac{f}{100\mbox{Hz}}\right]\right) 
= \tau_r^{-1} \left[\frac{M}{M_{\odot}}\right]\left[\frac{f}{100\mbox{Hz}}\right] \delta\nonumber\\
&+& 2.0\times 10^{-6}\left[\frac{\dot{M}}{1 M_{\odot}\mbox{y}^{-1}}\right] \left[\frac{M}{M_{\odot}}\right]^{1/2}.
\label{eq-freq}
\end{eqnarray}

\subsection{Terminal values of the temperature and the rotational frequency of 
nearly constant mass models }

It is important to note that the thermal balance as well as the angular 
momentum balance determines the approximate terminal values of the temperature 
and the rotational frequency of the star, if the accretion rate is fixed and 
the mass is regarded as nearly constant in the evolution.

By setting the right hand side of Eq.~(\ref{eq-thermal}) to be zero, the 
temperature can be solved as,
\begin{equation}
	T_9 = 3.0~\left[\frac{\dot{M}}{1 M_{\odot} \mbox{y}^{-1}}\right]^{1/8} \left[\frac{M}{M_{\odot}}\right]^{7/48} \ .
\label{t9expr}
\end{equation}
Thus for the star with mass $M\sim 1.4 M_{\odot}$ and the accretion rate with
$\dot{M}\sim 1 M_{\odot} \mbox{y}^{-1}$, the (uniform) temperature will be 
about $3.5\times 10^9$K, which is near the most susceptible temperature for
the r-mode instability to grow~\cite{lindblom-owen-morsink}.

The equation of the angular momentum balance gives the terminal value
of the rotational frequency of the star by setting the right hand side of 
Eq.(\ref{eq-freq}) to be zero. If we omit the contributions of viscous 
damping whose effect is smallest near at the temperature obtained above, we
get the expression for the terminal rotational frequency as :
\footnote{After the submission, we find a paper on an effect of a magnetic field to the r-mode instability\cite{rezzolla-et-al}. Presence of the magnetic field may weaken the instability significantly. If this is the case with the situation of our interest here, the terminal frequency will be much higher.}
\begin{equation}
	\left[\frac{f}{100\mbox{Hz}}\right] = 1.3~\left[\frac{\dot{M}}{1 M_{\odot} \mbox{y}^{-1}}\right]^{1/7} \left[\frac{M}{M_\odot}\right]^{-3/14} \delta^{-1/7}.	
\label{f100expr}
\end{equation}

\subsection{Evolutionary curves}

Given the mass accretion rate and the initial values of the stellar mass, 
temperature and frequency, the evolution of the corresponding quantities
can be followed by integrating Eqs.~(\ref{eq-mass}), (\ref{eq-thermal})
and (\ref{eq-freq}). We have computed several evolutionary sequences by 
choosing different sets for the initial values.

\begin{figure}
        \centering\leavevmode
        \psfig{file=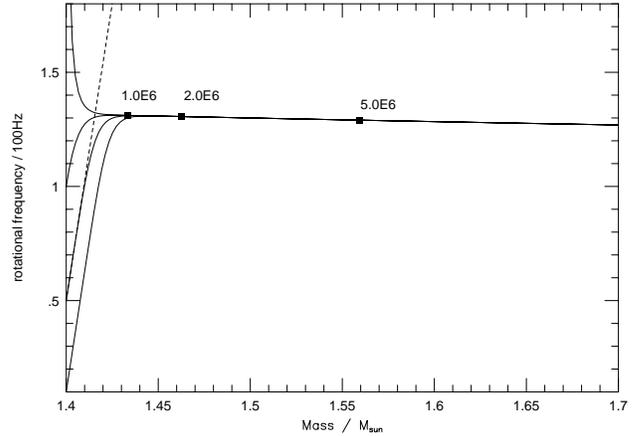,width=9.5cm,angle=90,clip=}
\caption{Evolutionary curves of the rotational frequency 
$f/100\mbox{Hz}$ for different values of the initial rotational frequency, 
i.e. $f/100\mbox{Hz} = 0.1, 0.5, 1$ and $3$, are plotted against the stellar 
mass which can be regarded as the `time' variable. Model parameters
are as follows: the initial mass $M=1.4M_{\odot}$, the initial temperature 
$T_9=10^{-3}$, the mass accretion rate $\dot{M} = 1 M_{\odot}\mbox{y}^{-1}$. 
Numbers attached to the curves are the values of the lapse of time in 
units of second. Also plotted is the curve of the model in which 
the r-mode instability is not included (dashed curve). 
}
\label{fig1}
\end{figure}

\begin{figure}
        \centering\leavevmode
        \psfig{file=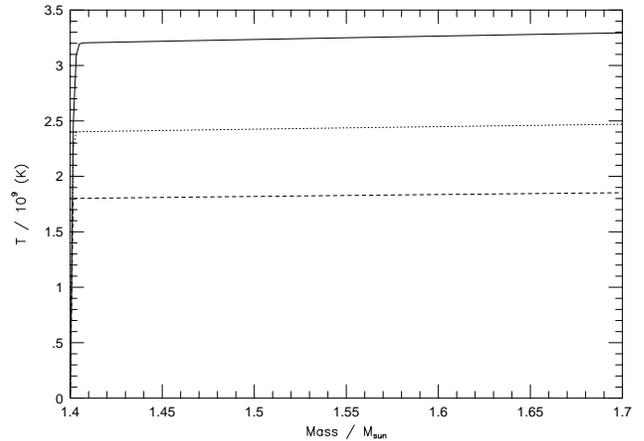,width=9.5cm,angle=90,clip=}
\caption{Evolutionary curves of the temperature $T_9=T/10^9$(K) 
for different values of the mass accretion rate, i.e. 
$\dot{M}/(1 M_{\odot}\mbox{y}^{-1}) = 1$ (solid curve), $10^{-1}$ 
(dotted curve), $10^{-2}$ (dashed curve), are plotted against the 
stellar mass. Model parameters are as follows:
the initial mass $M=1.4M_{\odot}$, the initial temperature $T_9=10^{-3}$, 
and the initial rotational frequency $f/100\mbox{Hz}=0.5$.
}
\label{fig2}
\end{figure}

We will show how the evolution depends on the initial values.
In Fig.~\ref{fig1} the rotational frequency is plotted against the mass of
the star which serves as the `time' variable because the mass accretion rate
is assumed to be constant. With the mass accretion rate fixed, solutions with
different values of the initial frequency converges to a certain value 
rather quickly compared with the mass accretion time scale. This behaviour is
expected from the rapid temperature saturation seen in Fig.~\ref{fig2}
by swift establishment of the thermal balance between heating and cooling.
It should be pointed out that the omission of the r-mode instability
would lead to very rapid increase of the rotational frequency and
that the star would reach its mass-shedding limit before the accretion 
induced collapse occurs.
%
\begin{figure}
        \centering\leavevmode
        \psfig{file=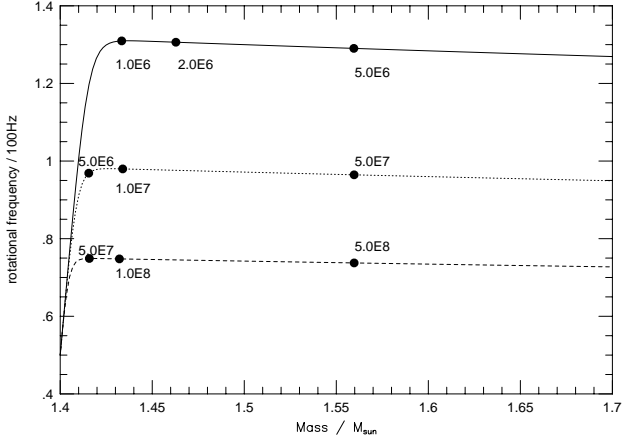,width=9.5cm,angle=90,clip=}
\caption{The rotational frequency is plotted against the stellar
mass. Three lines correspond to the three temperature lines in Fig.~2
($\dot{M}/(1 M_{\odot}\mbox{y}^{-1}) = 1$ (solid curve), $10^{-1}$ 
(dotted curve), $10^{-2}$ (dashed curve)).
Numbers attached to these lines are the values of the lapse of time
in units of second.
}
\label{fig3}
\end{figure}

The mass accretion rate dependence of the frequency evolution is 
shown in Fig.~\ref{fig3}. As seen in Fig.~\ref{fig2} the temperature
of the star quickly saturates at the limiting value which depends on the 
mass accretion rate. Since the strength of the r-mode instability depends
on the temperature as well as the stellar rotational frequency, the 
rotational frequency also quickly saturates. As the time scale of 
the angular momentum change is longer than that of the temperature change,
the frequency saturation is reached slightly later compared with 
the temperature saturation.

\section{Discussions}

\subsection{Formation of slowly rotating black holes}

We have shown that the r-mode instability sets a severe limit on
the rotational frequency of the hypercritically accreting neutron stars 
proposed by Bethe et al.~\shortcite{bethe-brown-lee}. As the mass accretion 
rate dependence of the rotational frequency (Eq.~(\ref{f100expr})) is rather 
weak, the rotational frequency of the star is well below its mass-shedding 
limit. Even if the mass accretion rate rises to $\dot{M}=10^2 
M_{\odot}\mbox{y}^{-1}$, the rotational frequency increases only to $200$Hz. 
Thus the gravitational collapse of neutron stars due to mass accretion to low 
mass black holes in these systems will proceed almost spherically and
black holes formed just after the collapse will rotate so slowly
that the rotational parameter is not large either, i.e. $q=cJ/GM^2 \sim 0.01$ 
for $1.7 M_{\odot}$.

\subsection{Observational possibility of gravitational waves from the 
hypercritically accreting neutron stars}

The amount of gravitational radiation from the system can be evaluated
as follows. The process discussed in this paper is similar to that 
studied by Wagoner~\shortcite{wagoner}. A neutron star in an X-ray binary
system emits gravitational wave so as to balance the accretion torque with
the gravitational radiation torque of the f-mode (`Wagoner star'). 
For our models the f-mode must be replaced by the r-mode and the accretion 
rate is considerably different, i.e. increased. The standard formula of 
the luminosity of gravitational radiation can be expressed 
as~\cite{shapiro-teukolsky},
\begin{equation}
\frac{c^3}{G}\omega^2h^2 D^2 \sim 2\pi f \dot{M} j_a \ ,
\end{equation}
where $\omega, h, D$ are the eigenfrequency of the mode, the 
dimensionless strain of the gravitational field and the distance 
from the source, respectively. 
Using the terminal rotational frequency $f$ obtained above
and the relation of the eigenfrequency and the angular velocity
for $l = m = 2$ modes,
\begin{equation}
 \omega = \frac{4}{3}\cdot 2\pi f \,
\end{equation}
we get
\begin{equation}
	h \sim 5.6\times 10^{-22} \left[\frac{\dot{M}}{1 M_\odot\mbox{y}^{-1}}\right]^{3/7}\left[\frac{M}{M_\odot}\right]^{5/14}\left[\frac{D}{1\mbox{kpc}}\right]^{-1} \delta^{1/14}.
\end{equation}
As seen from the result obtained in this paper, the rotational frequency 
becomes almost constant after a short time of accretion. Thus it is rather
easy to detect signals of the gravitational wave by integrating it for 
an appropriate interval of time. 

We here define the Fourier component of the squared dimensionless strain $h^2$
as $\hat{h}^2$. Then from the Parseval's relation,
\begin{equation}
	\hat{h}^2\cdot \sqrt{\Delta\omega} \sim h^2\cdot \sqrt{\frac{4\pi}{\omega}N_{\rm cycle}},
\end{equation}
where $N_{\rm cycle}$ is the number of cycles of the oscillation 
in the interval of observation, $\Delta t_{\rm obs}$. 
Frequency width $\Delta\omega$ and $\Delta t_{\rm obs}$ are related by the
uncertainty principle as,
\begin{equation}
	\Delta\omega\cdot\Delta t_{\rm obs} \ge \frac{1}{2}.
\end{equation}
As $\Delta t_{\rm obs} = \frac{4\pi}{\omega}N_{\rm cycle}$, the r.m.s. of the
Fourier component of the strain is,
\begin{equation}
	\hat{h} \le \frac{\sqrt{\Delta t_{\rm obs}}}{2^{1/4}} h.
\end{equation} 

The event rate ${\cal R}$ of the hypercritical accretion is quite uncertain. 
The maximum value is expected to be 
$\sim 10^{-4}\mbox{y}^{-1}\mbox{galaxy}^{-1}$, which is the birth rate
of the close black hole--neutron star binary~\cite{bethe-brown}.
In order to detect $N_0$ events per year, we have to observe the region
with the distance as follows:
\begin{equation}
D=(3N_0/4\pi{\cal R}n_{\rm gal})^{1/3} \ ,
\end{equation}
where $n_{\rm gal} \sim 0.01\mbox{Mpc}^{-3}$~\cite{efstathiou}. 
Here $n_{\rm gal}$ is the number density of the massive galaxies 
($\sim 10^{11}M_{\odot}$). 

Then Fourier component of the strain is,
\begin{eqnarray}
	&&\hat{h}(\mbox{Hz}^{-1/2}) \la 1\times 10^{-23} \cdot\delta^{1/14}\left[\frac{\dot{M}}{1M_\odot \mbox{y}^{-1}}\right]^{3/7}\\\nonumber
&&\times\left[\frac{M}{M_\odot}\right]^{5/14}\left[\frac{{\cal R}}{10^{-4}\mbox{galaxy}^{-1}\mbox{y}^{-1}}\right]^{1/3} N_0^{-1/3}\left[\frac{t_{\rm obs}}{1~{\rm month}}\right]^{1/2}
\end{eqnarray}
Comparing this with the expected 
sensitivity of the {\it Laser Interferometer Gravitational-Wave 
Observatory}~(LIGO),  the integration of one month is sufficient for the 
detection by the LIGO II facility ~(see Shoemaker 2000 for the sensitivity 
curves of the LIGO). As the dependence of the strain on ${\cal R}$ is weak,
the LIGO II can see the signal by the integration of one month, even if 
${\cal R}$ is reduced by two orders of magnitude.
Typical frequency of the source may be around $100\sim 200$Hz.

\subsection{Uncertainties in our analysis}

In this investigation we have applied the `one zone' approximation for 
simplicity. It means that the variables such as the stellar temperature 
or the rotational frequency are represented by some `averaged' values
throughout the whole star. It premises that the viscosity works very
efficiently compared to the time scale of the evolution of the system,
typically about one year.
The adjustment time scale of shear viscosity to smooth 
out the differential flow can be estimated for the neutron-neutron 
scattering dominant matter~\cite{lindblom-owen-morsink} as follows:
\begin{equation}
	\tau_{\rm adj} \la \frac{R^2 \bar{\rho}}{\eta} \sim \frac{R^2 T^2}{347 \bar{\rho}^{5/4}}.
\end{equation}
Assuming the stellar radius $R=10$km and the averaged density 
$\bar{\rho}=3\times 10^{14}\mbox{g/cm}^3$, this amounts to $0.8$ y for 
$T=10^8$K. The rotation law of the star can 
deviate from the uniform rotation if this time scale is comparable or
longer than the spin-up time.
In that case, behaviour of (classical) 
r-modes is uncertain and their characteristics may be drastically changed. 
We here {\it assume} that they 
can be approximated by some kind of average of the rotational angular velocity.
The validity of this assumption should be confirmed elsewhere.

It has been pointed out that in the hypercritical accretion process, 
explosive outflows driven by neutrino heating~\cite{fryer-benz-herant} 
or jet formation~\cite{armitage-livio} may significantly reduce the mass 
attachment onto the neutron star. If it is the case, it is more 
probable that double neutron star binary systems will be formed through 
this channel. In such accretion inefficient situations, the mass accretion 
rate $\dot{M}_{\rm acc}$ is lower than the mass inflow rate $\dot{M}_{\rm in}$ 
from the envelope of the companion star. If the accretion inefficient flows 
behave similarly as the accretion efficient flows, evolution of the stars may 
be approximated by our models with lower $\dot{M}$ and with a 
shorter duration, for the same value of $\dot{M}_{\rm in}$. This results in 
smaller terminal rotational frequencies of the star. Thus neutron stars which 
survive after mass accretion in this process would have even lower rotational 
frequencies.\footnote{It is possible that the characteristics of the accretion
flow change drastically and that the accreted matter carries less angular 
momentum than the case considered here. Of course this will lead
to the even more reduction of the terminal rotational frequency.}

It is a surprise that the possibility of observation of gravitational waves
from the systems seems unexpectedly large. This might encourage the ongoing 
detection experiments of gravitational waves. However we have to be careful 
in concluding the significance of the hypercritically accreting neutron stars 
as sources of gravitational waves. Our analysis here might be too simplified 
and naive to get conclusive answers to the observational possibility. 
Some simplified assumptions may influence the result significantly. 
For instance, more realistic behaviour of the accretion flow and the accreted
matter may drastically change the efficiency of the accretion torque.
Angular momentum re-distribution in the realistic star may modify 
the r-mode characteristics significantly, which may affect
the strength of the instability. Thus the uncertainty included in
our assumptions must be investigated more intensively in the future.

\section*{Acknowledgment}
We would like to express our gratitude to Dr. N. Andersson, the referee,
for his suggestion that the detectability of the gravitational wave signals 
may be rather high.
%

%

\begin{thebibliography}{}
%
	\bibitem[\protect\citename{Andersson }1998]{andersson}
		Andersson N., 1998, ApJ, 502, 708 
	\bibitem[\protect\citename{Andersson et al. }1999]{andersson-kokkotas-stergioulas}
		Andersson N., Kokkotas K., Stergioulas N., 1999, ApJ, 516, 307 
	\bibitem[\protect\citename{Armitage \& Livio }1999]{armitage-livio}
		Armitage P. J., Livio M., 1999,
		ApJ submitted~(astro-ph/9906028)
	\bibitem[\protect\citename{Bethe \& Brown }1998]{bethe-brown}
		Bethe H. A., Brown G. E., 1998, ApJ, 506, 780 
	\bibitem[\protect\citename{Bethe et al. }1999]{bethe-brown-lee}
		Bethe H. A., Brown G. E., Lee C.-H., 1999, 
		ApJ submitted~(astro-ph/9909132)
	\bibitem[\protect\citename{Brown }1995]{brown}
		Brown G. E., 1995, ApJ, 440, 270 
	\bibitem[\protect\citename{Chevalier }1993]{chevalier93}
		Chevalier R. A., 1993, ApJ, 411, L33 
	\bibitem[\protect\citename{Chevalier }1996]{chevalier96}
		Chevalier R. A., 1996, ApJ, 459, 322 
	\bibitem[\protect\citename{Efstathiou et al. }1988]{efstathiou}
		Efstathiou G., Ellis R. S., Peterson B., 1988, MNRAS, 232, 431
	\bibitem[\protect\citename{Friedman \& Morsink }1998]{friedman-morsink}
		Friedman J. L., Morsink S. M., 1998, ApJ, 502, 714 
	\bibitem[\protect\citename{Fryer et al. }1998]{fryer-benz-herant}
		Fryer C. L., Benz W., Herant M., 1996, ApJ, 460, 801 
	\bibitem[\protect\citename{Levin }1999]{levin}
		Levin Y., 1999, ApJ, 517, 328 
	\bibitem[\protect\citename{Lindblom et al. }1998]{lindblom-owen-morsink}
		Lindblom L., Owen B. J., Morsink S. M., 1998, Phys. Rev. Lett., 80, 4843 
	\bibitem[\protect\citename{Narayan \& Yi }1994]{narayan-yi}
		Narayan R., Yi I., 1994, ApJ, 428, L13 
	\bibitem[\protect\citename{Owen et al. }1998]{owen-et-al}
		Owen B. J., Lindblom L., Cutler C., Schutz B. F., Vecchio A., Andersson N., 1998, Phys. Rev. D58, 084020
	\bibitem[\protect\citename{Rezzolla et al. }2000]{rezzolla-et-al}
		Rezzolla L., Lamb F. K., Shapiro S. L., 2000, ApJ, 531, L139
	\bibitem[\protect\citename{Shapiro \& Teukolsky }1983]{shapiro-teukolsky}
		Shapiro S. L., Teukolsky S., 1983, {\it Black Holes, White Dwarfs and Neutron Stars}, Wiley, New York
	\bibitem[\protect\citename{Shoemaker }2000]{shoemaker}
		Shoemaker D., 2000, in {\it Matters of Gravity, No.15} (http://gravity.phys.psu.edu/mog/mog15/node14.html)
	\bibitem[\protect\citename{Thorne }1980]{thorne}
		Thorne K. S., 1980, Rev. Mod. Phys, 52, 299 
	\bibitem[\protect\citename{Wagoner }1984]{wagoner}
		Wagoner R. V., 1984, ApJ, 278, 345 

\end{thebibliography}
\end{document}